# Closely spaced and separately contacted 2d electron and 2d hole gases by *in situ* focussed ion implantation

M. Pohlt, M. Lynass[a], J.G.S. Lok, W. Dietsche[b], K. v. Klitzing and K. Eberl

Max-Planck-Institut für Festkörperforschung, Heisenbergstrasse 1, 70569 Stuttgart, Germany

R. Mühle

Swiss Federal Institute of Technology Zürich, ETH Zentrum, 8092 Zürich, Switzerland

Abstract

Separately contacted double layers of a 2d electron - 2d hole gases have been prepared in GaAs separated by thin AlGaAs barriers with thicknesses down to 15 nm. The molecular-beam-epitaxial growth was interrupted just before the barrier in order to use *in situ* focussed-ion-beam implantation to pattern contacts which extend underneath the barrier. The two charge gases form upon biasing the p- and n-type contacts underneath and above the barrier in forward direction and show independent transistor-like behavior.

[a] also at University of Bath, Bath, United Kingdom

[b] E-Mail: dietsche@klizix.mpi-stuttgart.mpg.de

There has been great interest over the last years in fabricating closely spaced 2d electron and 2d hole systems by molecular beam epitaxy (MBE). If the distance between the oppositely charged layers can be decreased to about 10 to 20 nm then one should be able to observe new correlation phenomena like bound states of spatially separated electrons and holes. A gas of such indirect excitons could even condense into a superfluid as was proposed by several authors [1].

The crucial point in preparing such systems is the small distance between the two charge layers which makes their preparation impossible if one relies solely on the doping of conventional semiconductors because of the high built-in electric field. Several routes have been followed to circumvent this problem. One possibility is the use of the InAs-GaSb heterostructures which have a large band offset leading to equilibrium 2d electron (2DEG) and 2d hole gases (2DHG) in close proximity [2]. However, many proposed experiments require separate contacts to the two layers which, to our knowledge, have not yet been realised in this system. Furthermore, it seems to be difficult to adjust the densities of the two gases by external gates and to obtain large mean free paths of the holes and electrons [3].

Another rather successful route to closely spaced layers is the optical excitation of carriers in quantum wells. The application of an electric field separates the electrons and holes. Interaction effects between the two are usually studied using the luminescence and indications of the existence of Bose-Einstein condensation have indeed been observed [4]. However, the lack of electrical contacts to the optically excited carriers make transport measurements nearly impossible. Furthermore, only systems where the recombination life time is sufficiently short to cause sufficient luminescence can be studied this way.

An alternative is the use of nonequilibrium methods which means that a voltage is applied be-tween the two layers compensating the band bending between the positive and negative charges. These techniques rely on separate contacts to the two layers. The contacts must be nearly perfect because the charge gases themselves form only after an electric field has been applied. Thus, the contacts should overlap with the barrier region or should at least be perfectly aligned. Several nonequilibrium designs have been published. The contacts to the bottom charge layer were either done by ion implantation using the gate as a self-aligned mask [5] or by evaporation from the side of the structure and subsequent diffusion [6]. Another group used ion implantation during an MBE growth interruption but did not succeed to produce high-quality structures [7]. In all three experiments it was difficult to obtain good contacts to the bottom layer without causing unacceptable leakage currents across the barrier.

In this publication, we present a reliable technology for preparing spatially separated 2DEG and 2DHG systems at barrier thicknesses down to 15 nm which is also based on the nonequilibrium method. A growth interruption in the MBE process is used to transfer the sample into a focussed-ion beam system (FIB) where the contacts to the lower charge gas are defined by ion implantation prior to the growth of the barrier and the remaining layers.

Our device structure is shown schematically in Fig 1 (a). The 2DEG and the 2DHG form on top and below the AlGaAs barrier (marked in black color) by the application of a voltage between the electron contacts on the top of the structure and the hole contacts below the barrier. The extension of p-doped contact region underneath the barrier is the crucial point of this design and is achieved by the FIB implantation [8]. An n-type gate at the very bottom of the structure and a metallic front gate allow the individual adjustment of the two carrier concentrations. The following growth and implantation sequence turned out to be successful to prepare the required structures:

- Semi insulating GaAs-substrate
- 500 nm GaAs buffer layer
- 100 nm Si-doped ($2x10^{18}$ $cm^{-3}$) GaAs as the backgate layer
- 300 nm undoped GaAs
- 100 nm C-doped ($2x10^{18}$ $cm^{-3}$) GaAs for p-contacts
- UHV-transfer into FIB-chamber
- $Si^+$ implantation ($10^{12}$ $cm^{-2}$ at 30keV) to isolate part of the p-layer
- UHV-transfer back to MBE-chamber
- 300 nm GaAs with C-doping gradually decreasing from $10^{17}$ $cm^{-3}$ to $5x10^{16}$ $cm^{-3}$
- 50 nm undoped GaAs
- 15 nm $Al_xGa_{1-x}As$ barrier with x=0.8
- 50 nm undoped GaAs
- 300 nm GaAs with Si-doping gradually increasing from $10^{16}$ $cm^{-3}$ to $10^{17}$ $cm^{-3}$
- 200 nm GaAs with Si-doping of $2x10^{18}$ $cm^{-3}$

Strong efforts had to be made to avoid contamination during the transfer process to and from the FIB chamber. The pressure in the transfer tunnel is typically $10^{-9}$ mbar which causes contamination with carbon and oxygen as revealed by SIMS analysis. The oxygen can be removed after the return to the MBE by a heat treatment at 630° C for 15 minutes. The carbon could only be removed by depositing an $As_2$ layer before the transfer to the FIB. This was done using a GaAs cracker cell at 830° C for 30 sec with the substrate at room temperature. After return to the main chamber the $As_2$ cap evaporates together with the contaminants at temperatures above 350 °C. Using this procedure, SIMS data did not show residual C contamination.

The Si ions easily penetrate the $As_2$ layer and overcompensate the C-doping of the p-contact layer. Alternatively, one could have used a p-type implantation, e.g. Be ions to form the p-type contacts. This was tested but was not successful due to the Be segregation with further growth leading to leakage through the barrier. For the alignment of the implanted regions with the subsequent photolithography we implant Au marks with high doses that are easily visible after overgrowth.

After the completion of the MBE growth the samples are annealed at 800°C for 5 sec in order to activate the implanted Si donors. Then the devices are prepared conventionally. First, the p-type contacts are made with alloyed Au(5nm)/Zn(30nm)/Au(25nm) after etching most of the material atop the p-contact regions. Then the shallow n-type contacts are made by annealing very thin Au/Ge/Ni layers of only 4 nm total thickness at 320 °C for 30 s. This way the n-type contacts at the top are very shallow and do not diffuse into the barrier. The fabrication of the self aligned front gate is performed in two steps with one lithography: first the 200 nm strongly doped GaAs layer is etched away separating the n-contacts from each other. Then a Ti/Au metallization is evaporated with the same mask under rotation and a tilt angle of about 20°. This procedure minimises the lateral distance between the gates and the n-contacts. In a last step the mesa is defined with reactive ion etching which prevents any metallization from the contacts or the front gate from making an unintentional contact with the region under the barrier. This is crucial because the Schottky barriers provide no sufficient isolation to the relatively high voltage between 2DEG and 2DHG. A photograph of the completed structure is shown in Fig 1 (b).

The samples obtained this way contain two channels which are populated by applying a voltage between the front- and the backgate with the source contacts of both channels connected to the respective gate. The output characteristics of the p and n channels are shown in Fig 2. The barrier of this sample is 15.3 nm $Al_{0.8}Ga_{0.2}As$.

The contacts to the two channels are completely ohmic as is visible in Fig. 2 from the initially linear increase of the currents with the source-drain voltage. No conductivity is measurable at voltages below the respective threshold voltage indicating that there are no parallel conductance channels. Each of the two charge gases behaves like an enhancement-mode transistor. Magnetotransport measurements of the two charge gases (not shown) exhibit the typical Shubnikov-de-Haas oscillations in the longitudinal resistance proving their two dimensional nature and give mobilities of 130,000 $cm^2$/Vs and of 47,000 $cm^2$/Vsec for the 2DEG and the 2DHG system respectively at densities of about $3x10^{11}$ $cm^{-2}$. These mobilities are about half of those obtained in the standard heterostructures from the same MBE system containing just one charge gas without growth interruption and FIB transfer. Leakage current is about 300 pA/$mm^2$ at 1.6 V. This corresponds to a tunneling lifetime of 0.4 sec which is at least 6 orders of magnitude longer than observed with 2DEG/2DHG systems generated by optical excitation.

The threshold voltage is 1.5 V for the n-channel and 1.56 V for the p-channel of the sample in Fig. 2. These slightly different thresholds originate probably from different charging of the two gates and the respective doped regions between them and the actual active layers. This can be avoided by slightly different structures which are of the SIS-FET (semiconductor-insulator-semiconductor) type. The SIS-FETs do not have a metallic front gate but just one contact that connects to the entire upper 2DEG. The difference in the threshold behavior of the two devices is visible from Fig. 3. The dashed line shows that the charges form in two steps with increasing voltage between the two layers if a separate front gate is present. The solid line is measured with a SISFET; both charge layers form in a single step. The realisation of small but identical densi-ties in the two layers is very important because such conditions are preferable for the observation of excitonic effects. Further capacitance data in this regime will be published in a future paper [9].

In conclusion, we have prepared spatially separated 2d electron and 2d hole layers with distances down to 15 nm using FIB implantation during an MBE growth interruption. Carrier concentrations in the two layers can be varied from $10^{10}$ $cm^{-2}$ up to nearly $10^{12}$ $cm^{-2}$. Both transport and capacitance measurements are possible with these devices and will be used to study the interaction effects between the layers.

We thank U. Sivan for helpful discussion and W. Wegscheider for the growth of several wafers in the course of this work. The contributions of H. Rubel and C. Jörger at the early stage of the project are gratefully acknowledged. This research is supported by the BMBF.

# References

[1] P.B. Littlewood and X. Zhu, Physica Scripta **T68**, 56 (1996) and references therein.

[2] E.E. Mendez, L. Esaki, and L.L. Chang, Phys. Rev. Lett. **55**, 2216 (1985)

[3] T. P. Marlow, L. J. Cooper, D. D. Arnone, N. K. Patel, D. M. Whittaker, E. H. Linfield, D. A. Ritchie, and M. Pepper , Phys. Rev. Lett **82**, 2362 (1999)

[4] T. Fukuzawa, E. E. Mendez and J. M. Hong, Phys. Rev. Lett. **64**, 3066 (1990); J. A. Kash, M. Zachau, E.E. Mendez, J. M.Hong and T. Fukuzawa, Phys. Rev. Lett. **66**, 2247 (1991); L.V. Butov, A. Zrenner, G. Abstreiter, G. Böhm and G. Weimann, Phys. Rev. Lett. **73**, 304 (1994)

[5] U. Sivan, P.M. Solomon and H. Shtrikman, Phys. Rev. Lett. 68, 1196 (1992)

[6] B.E. Kane, J.P. Eisenstein, W. Wegscheider L.N. Pfeiffer and K.W. West, Appl.Phys.Lett. 65, 3266 (1996)

[7] S. Vijendran, P.J.A. Sazio, H.E. Beere, G.A.C. Jones, D.A. Ritchie, and C.E: Norman, J. Vac.Sci.Tech. **B17**, 3226 (1999)

[8] Orsay Physics FIB column with integrated optical microscope.

[9] M. Pohlt, J.G.S.Lok, M. Lynass et al. to be published.

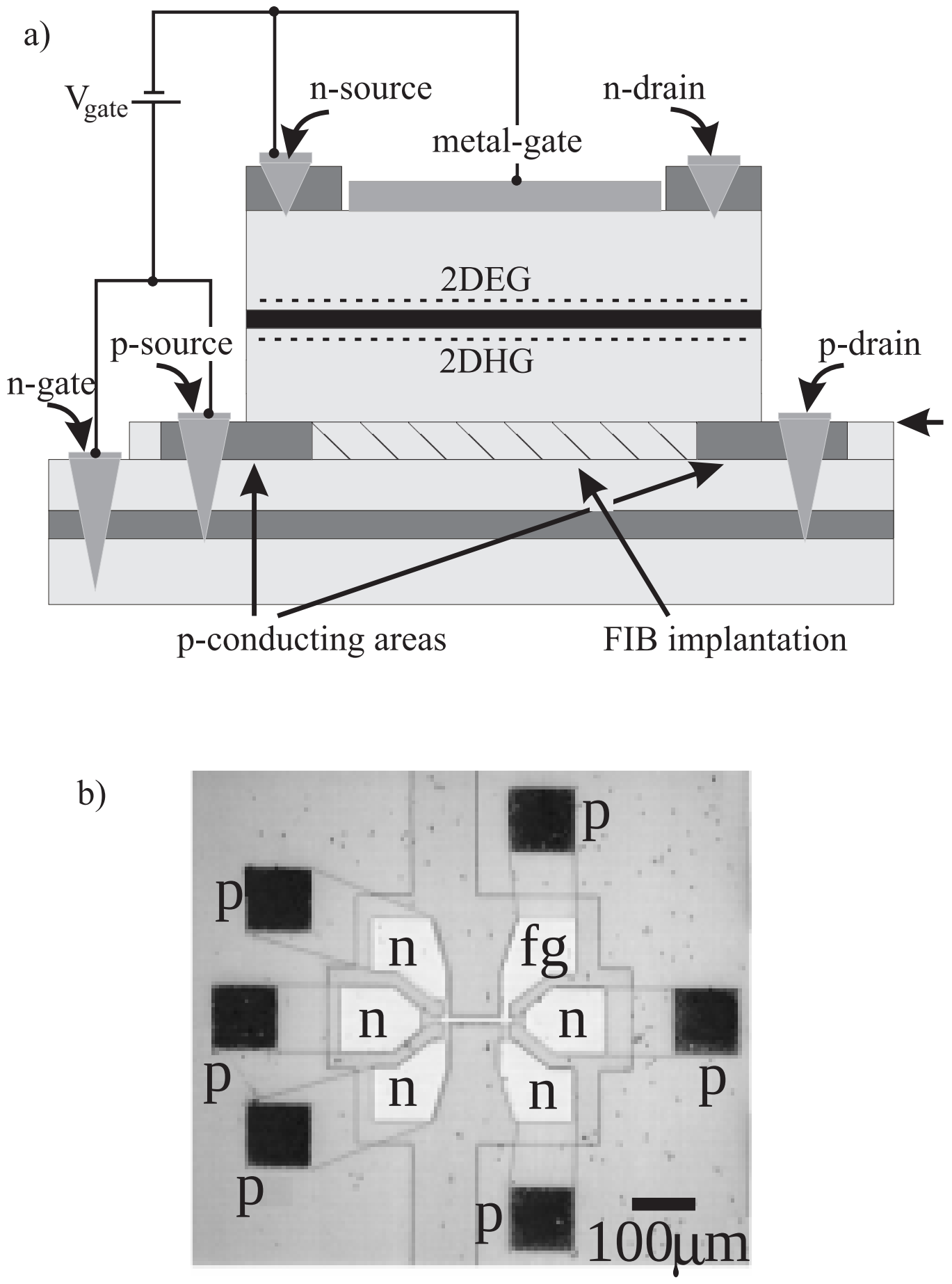

Fig. 1

(a) Sample structure. The 2DEG and the 2DHG form on the upper and the lower side of the AlGaAs barrier (black) after a voltage $V_{gate}$ has been applied between the p and n contacts. The p-contacts are structured by Si- implantation (hatched area) using a FIB system in a MBE growth interruption. The horizontal arrow marks the layer of the growth interruption.

(b) Photograph of a device allowing separate transport measurements of the 2DEG and the 2DHG. „p“ and „n“ mark the p- and n-type contacts. „fg“ connects to the front gate. The area structure extending from the top to the bottom of the picture has been etched down to the p-type contact layer.  The contact to the backgate is not visible in this picture.

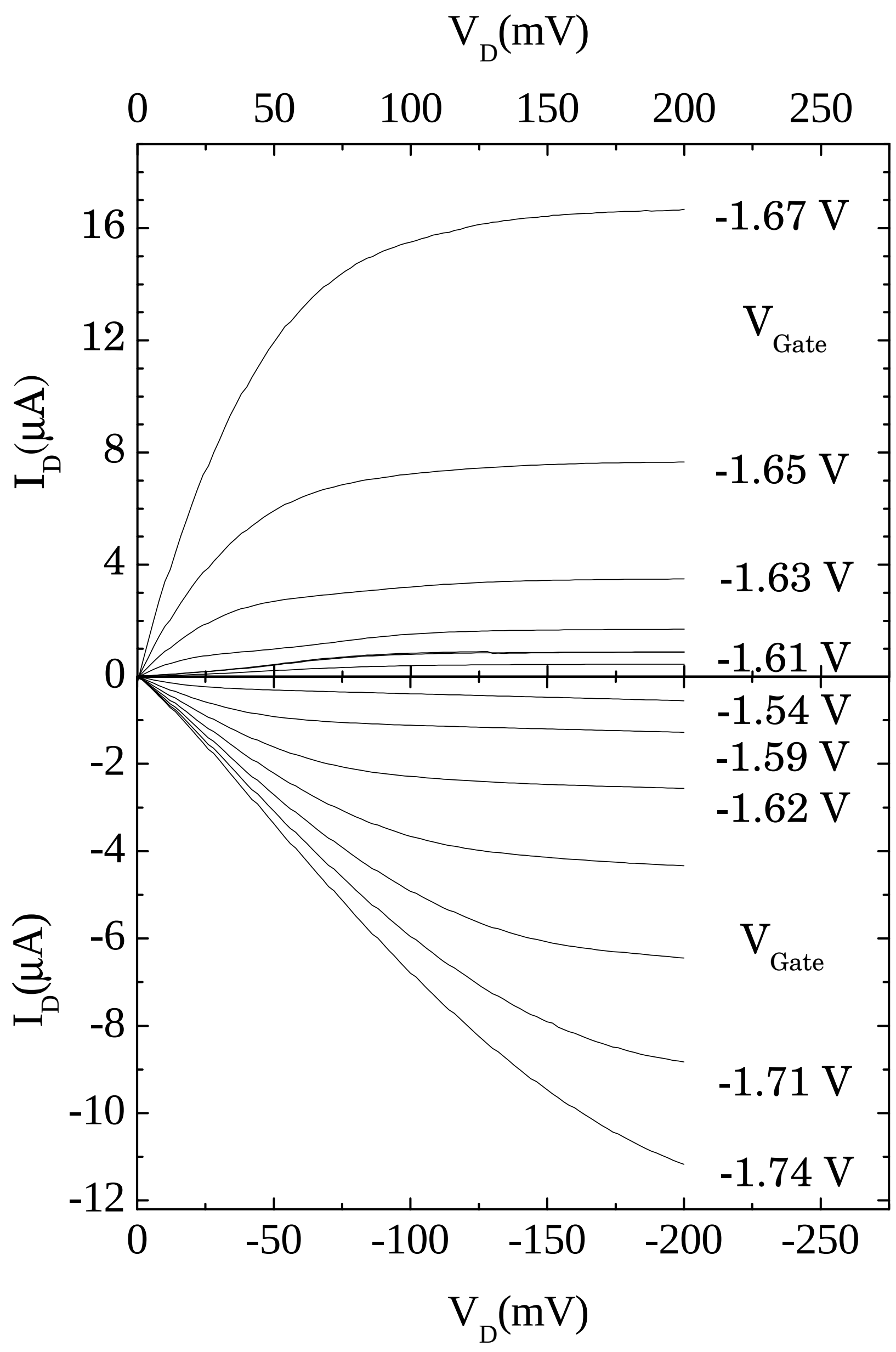

Fig. 2

Transistor characteristics of the 2DEG and the 2DHG forming at the two sides of an Al-GaAs barrier with 15 nm thickness at 4.2 K.

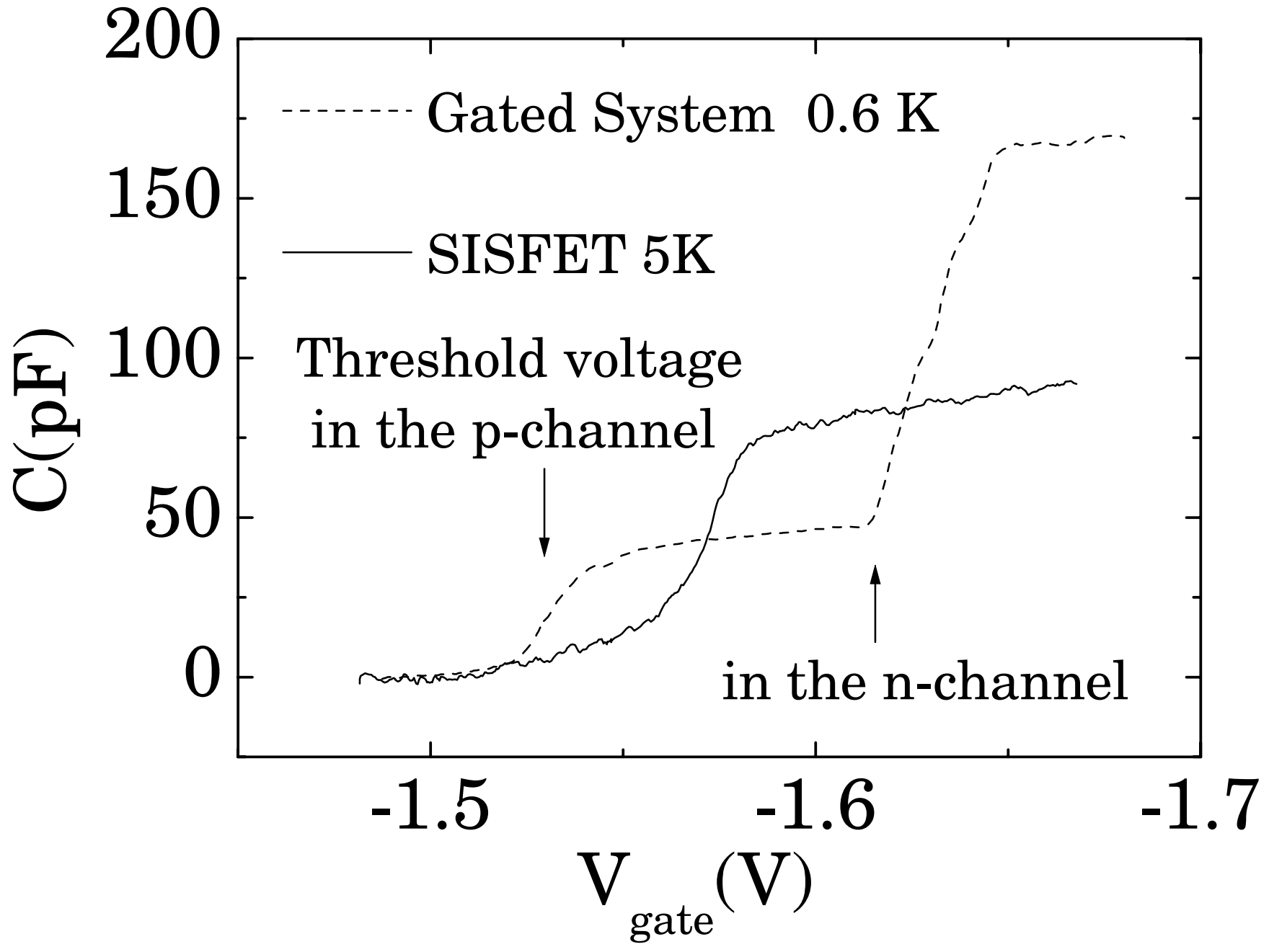

Fig. 3

Capacitance measurements of the device of Fig. 2 (dashed line) showing the onset behavior of the two charge gases at two different theshold voltages. In a SIS FET structure the onset occurs simultaneously. The barriers of these samples are 20 nm each. A gate voltage increase of 0.1 V corresponds to an carrier density increase of about $2x10^{11}$ $cm^{-2}$.